\documentclass[12pt]{article}

\usepackage{latexsym,amsmath,amssymb,fancybox}

\usepackage[english]{babel}

\usepackage{color,epsfig}
\usepackage{graphicx}

\definecolor{navy}{rgb}{0.0,0.0,0.4}

\definecolor{rd}{rgb}{1,0,0}
\definecolor{or}{rgb}{1,.33,0}
\definecolor{pi}{rgb}{.66,.33,.33}
\definecolor{gn}{rgb}{0,.50,0}
\definecolor{be}{rgb}{0,0,.66}
\definecolor{ru}{rgb}{.66,0,.33}
\definecolor{vi}{rgb}{.33,0,.66}
\definecolor{gy}{rgb}{0,.33,.66}
\definecolor{ye}{rgb}{.66,.33,0}
\definecolor{bk}{rgb}{0,0,0}

\def\thf{\baselineskip=\normalbaselineskip\multiply\baselineskip
by 7\divide\baselineskip by 6}

\def\fff{\baselineskip=\normalbaselineskip}

\def\spose#1{\hbox to 0pt{#1\hss}}
\def\lta{\mathrel{\spose{\lower 3pt\hbox
{$\mathchar"218$}}\raise 2.0pt\hbox{$\mathchar"13C$}}}  \def\gta{\mathrel
{\spose{\lower 3pt\hbox{$\mathchar"218$}}\raise 2.0pt\hbox{$\mathchar"13E$}}}

\def\sqr#1#2{{\vcenter{\hrule height.4pt\hbox{\vrule width.8pt height#2pt
\kern#1pt\vrule width.8pt}\hrule height.4pt}}}

\def\spose#1{\hbox to 0pt{#1\hss}}\def\lta{\mathrel{\spose{\lower 3pt\hbox
{$\mathchar"218$}}\raise 2.0pt\hbox{$\mathchar"13C$}}}  \def\gta{\mathrel
{\spose{\lower 3pt\hbox{$\mathchar"218$}}\raise 2.0pt\hbox{$\mathchar"13E$}}}

\begin{document}

\def\be{\begin{equation}}
\def\fe{\end{equation}}

\newcommand{\eqn}{\label}
\newcommand{\bel}{\begin{equation}\label}

\def\eqdef{\fff\ \vbox{\hbox{$_{_{\rm def}}$} \hbox{$=$} }\ \thf }

\def\ov{\overline}


\def\Lr{ {\color{rd} {L}} }
\def\Jr{ {\color{rd} {J}} }
\def\calIr{ {\color{rd} {\cal I}} }
\def\Ar{ {\color{rd} {A}} }

\def\Br{ {\color{rd} {B}} }
\def\Cr{ {\color{rd} {C}} }
\def\Dr{ {\color{rd} {D}} }

\def\Xr{ {\color{rd} {X}} }\def\Yr{ {\color{rd} {Y}} }
\def\Rr{ {\color{rd} {R}} }
\def\kappar{ {\color{rd} {\kappa}} }

\def\calMr{ {\color{rd} {\cal M}} }

\def\nablar{ {\color{rd} {\nabla}} }
\def\deltar{ {\color{rd} {\delta}} }

\def\Tr{{\color{rd} T }}
\def\Pir{{\color{rd} {\mit\Pi} }}

\def\calXr{ {\color{rd} {\cal X}} }
\def\calUr{ {\color{rd} {\cal U}} }
\def\calTr{ {\color{rd} {\cal T}} }
\def\calVr{ {\color{rd} {\cal V}} }
\def\calPr{ {\color{rd} {\cal P}} }
\def\calBr{ {\color{rd} {\cal B}} }
\def\Gammar{ {\color{rd} {\Gamma}} }
\def\Psir{ {\color{rd} {\Psi}} }
\def\Sigmar{ {\color{rd} { X}} }
\def\Phir{ {\color{rd} {\Phi}} }
\def\phir{ {\color{rd} {\phi}} }
\def\varphir{ {\color{rd} {\varphi}} }
\def\Thetar{ {\color{rd} {\Theta}} }
\def\thetar{ {\color{rd} {\theta}} }

\def\grd{ {\color{rd} {g}} }
\def\wrd{ {\color{rd} {\mathfrak w}} }
\def\srd{ {\color{rd} {s}} }
\def\frd{ {\color{rd} {f}} }
\def\jrd{ {\color{rd} {j}} }
\def\erd{ {\color{rd} {e}} }
\def\egothrd{ {\color{rd} {\mathfrak e}} }

\def\dr{\spose {\raise 4.0pt \hbox{\color{rd}{\,\bf-}}} {\rm d}}


\def\Euro{{\color{ru}{\spose {\lower 2.5pt\hbox{${^=}$}}{\bf C}}}}

\def\Vru{ {\color{ru} {V}} }

\def\Gru{ {\color{ru} {G}} }
\def\kru{ {\color{ru} {k}} }

\def\calAr{ {\color{ru} {\cal A}} }
\def\calGr{ {\color{ru} {\cal G}} }
\def\calCr{ {\color{ru} {\cal C}} }
\def\ConStruc{ {\color{ru} {\copyright}} }
\def\omegaru{ {\color{ru} \omega}}
\def\Omegaru{ {\color{ru} \Omega}}
\def\alpharu{ {\color{ru} \alpha}}
\def\betaru{ {\color{ru} \beta}}
\def\gammaru{ {\color{ru} \gamma}}

\def\calDr{ {\color{ru} {\cal D}} }
\def\Dru{ {\color{ru} {D}} }
\def\aru{ {\color{ru} {a}} }
\def\Aru{ {\color{ru} {A}} }
\def\Fru{ {\color{ru} F} }
\def\amr{ {\color{ru}\bf{a}} }
\def\Amr{ {\color{ru}\bf{A}} }
\def\Fmr{ {\color{ru}\bf{F}} }
\def\wrru{ {\color{ru} {\wr}} }
\def\wru{ {\color{ru} {\vert\!\!\vert\!\!\vert}} }


\def\Libra{{\color{be}{\spose {\lower 2.5pt\hbox{${^=}$}}{\cal L}}}}

\def\gbe{{\color{be} g }}
\def\kbe{{\color{be} k }}
\def\sbe{{\color{be} s }}
\def\rhob{ {\color{be} {\rho}} }
\def\varpib{ {\color{be} {\varpi}} }
\def\vb{{\color{be} v }}
\def\partialb{ {\color{be} \partial}}
\def\nablab{ {\color{be} \nabla}}
\def\Gammab{ {\color{be} \Gamma}}
\def\Deltab{ {\color{be} \Delta}}
\def\Thetab{ {\color{be} {\Theta}} }
\def\Ab{{\color{be} A }}
\def\Rb{{\color{be} R}}
\def\db{\spose {\raise 4.0pt \hbox{\color{be}{\,\bf-}}} {\rm d}}
\def\Sigmab{ {\color{be} {\mit\Sigma}} }
\def\Sb{ {\color{be} S } }

\def\calSg{\ov{\color{gn}\cal S}}
\def\calS{{\color{gn}\cal S}}
\def\ggn{{\color{gn} g}}
\def\etag{{\color{gn}\eta}}
\def\deltag{{\color{gn}\delta}}
\def\nablag{ {\color{gn} \nabla}}
\def\Kg{{\color{gn} K}} 
\def\Gammag{{\color{gn} \Gamma}}
\def\perpg{{\color{gn}\perp\!}}
\def\xig{{\color{gn}\xi}}
\def\sigme{{\color{gy}\sigma}}

\def\gammar{ {\color{rd} {\gamma}} }
\def\psir{ {\color{rd} {\psi}} }
\def\chir{ {\color{rd} {\chi}} }
\def\Omegar{ {\color{rd} {\mit\Omega}} }
\def\nrd{ {\color{rd} {n}} }
\def\varrhob{ {\color{be} {\varrho}} }
\def\wgn{{\color{gn}\mathfrak w}}
\def\ugn{ {\color{gn} {u}} }
\def\ard{ {\color{rd} {a}} }

\def\brd{ {\color{rd} {b}} }

\def\hrd{ {\color{rd} {h}} }
\def\Prd{ {\color{rd} {P}} }
\def\prd{ {\color{rd} {p}} }
\def\rhor{ {\color{rd} {\rho}} }
\def\Ematr{ {\color{rd} {\mathfrak E}} }
\def\Er{ {\color{rd} {E}} }

\def\Agoru{ {\color{ru} {\mathfrak A}} }
\def\Fgoru{ {\color{ru} {\mathfrak F}} }

\def\crd{ {\color{rd} {c}} }

\def\mm{ {\color{or} {m}} }
\def\delth{ {\color{gn} {\delta}} }
\def\alpharo{ {\color{or} {\alpha}} }
\def\kapparo{ {\color{or} {\kappa}} }

\begin{center}
{\color{rd}\bf  FIELDS IN NONAFFINE BUNDLES. IV. \\[0.4cm]
Harmonious non-Abelian currents in string defects.}
\\[1cm]
 \underline{Brandon Carter} \\[0.6cm]
 \textcolor{ru}{LUTh (CNRS),
  Observatoire Paris - Meudon. }
  \\[0.5cm]
 {\color{be} 3 January, 2010.}
\end{center}  .

{\bf Abstract. } This article continues the study of the category 
of harmonious field models that was recently  introduced as a 
kinetically non-linear generalisation of the well known harmonic 
category of multiscalar fields over a supporting brane wordsheet 
in a target space with a curved Riemannian metric. Like the 
perfectly  harmonious case of which a familiar example is 
provided by ordinary barotropic perfect fluids, another important 
subcategory is the simply harmonious case, for which it is shown 
that as well as ``wiggle'' modes of the underlying brane world 
sheet, and sound type longitudinal modes, there will also be 
transverse shake modes that propagate at the speed of light. 
Models of this type are shown to arise from a non-Abelian 
generalisation of the Witten mechanism for conducting string 
formation by ordinary scalar fields with a suitable quartic self 
coupling term in the action.

\section{Introduction}
\label{Sec0}

This article continues a series in which a systematic covariant 
differentiation procedure was developed \cite{I,II} and applied
\cite{III} to multiscalar field models for which the relevant target 
space lacks the usual linear (vectorial or affine) structure, but is 
curved, as in the prototypical example \cite{Misner78} of a harmonic 
map, which is governed by dynamical equations involving non linear 
dependence on the fields but only linear dependence on their 
space-time gradients. As well as the possibility of confinement to 
string or brane worldsheets, and the gauging of internal symmetries 
of the target space, two different kinds of generalisation of the 
ordinary harmonic category were considered. The first \cite{II,III}
was that of forced-harmonic models in which a harmonic type kinetic 
term is supplemented in the Lagrangian by a self coupling term 
$\hat\calVr$ having the form of a predetermined scalar field on the 
target space. The second kind was that of harmonious models, whose 
definition will be recapitulated in the next section, and and which 
differ from ordinary harmonic and forced-harmonic models in that 
their dynamical equations involve nonlinear dependence not just on 
the fields but also on their gradients. 

Nonlinear gradient dependence of the kind in question has
long been familiar in the context of irrotational fluid models
such as are relevent for superfluidity \cite{CK92,C00} 
(and perhaps also for cosmology \cite{AMS01,LMR08}). However
the target space in these examples  is of the usual flat vectorial 
kind, as also is that of their cosmic string supported analogues
of both the singly \cite{C89,C95}  and multiply \cite{LMP09}
conducting kinds that have been studied in recent years.

Interest in such string supported fields began when Witten 
\cite{Witten85} drew attention to the fact that they would arise 
naturally by condensation in the cores of string (or other) 
topological defects in spacetime field models of the commonly 
considered -- kinematically linear -- kind in which the only 
nonlinearity is that of a scalar self coupling term responsible 
for spontaneous symmetry breaking of the vacuum. It was
subsequently recognised that suitable macroscopic models 
\cite{CP95,HC08} for the description of such fields in the thin 
string limit would need to involve nonlinear  field gradient 
dependence of the type qualifiable as harmonious, though only 
of the rather trivial kind in which the target space is flat 
and the relevant internal group Abelian

The main purpose of the present article is to show that 
a straightforward -- still kinematically linear -- extension
of the class of models proposed by Witten will give rise
to nonlinearly harmonious string models of a more interesting kind, 
in which the target space is curved and the relevant symmetry group
non Abelian.

Before proceeding it is important  to warn that the subject treated 
under the title ``non-Abelian string conductivity'' by Kibble and 
his collaborators \cite{KLY97} needs to be distinguished --
e.g.  by the insertion of another hyphen, so as to obtain
``non-Abelian-string conductivity'' -- from the subject of
of the present study, which might appropriately be entitled  
``non-Abelian string-conductivity''. The point of this nuance 
is that Kibble and his collaborators were concerned with a 
generalisation of Witten's model wherein, instead of being 
attributable to the spontaneous breakdown of an Abelian   
U$\{1\}$ symmetry, the string formation was attributable to the
spontaneous breakdown of a non-Abelian  SU$\{2\}$ symmetry,
but  the currents considered by these authors were nevertheless 
merely Abelian in the sense of being generated only by
a distinct U$\{1\}$ symmetry subalgebra that had survived the
breakdown. In contrast, the present work will be concerned with
a different kind of generalisation of Witten's model, wherein -- 
although the string itself will just be Abelian, in the sense of 
being attributable to the spontaneous breakdown only of an Abelian   
U$\{1\}$ symmetry subgroup -- the currents therein will actually 
be non-Abelian in the sense of being generated, not just by a  
U$\{1\}$ action, but by the action of a surviving non-Abelian 
SU$\{2\}$ or higher symmetry algebra.

Although the generalisation of the category of underlying 
kinetically linear spacetime field models is straightforward,
the Witten mechanism itself cannot be directly employed 
for the construction of a curved target space, as it depends
on a weak cylindrical symmetry ansatz that will not be self 
consistently applicable in the nonAbelian case. It will be 
shown below in Section \ref{Annex2} how
Witten's weak symmetry ansatz can be replaced for this purpose
by a more general local geodicity ansatz, that works as a good
approximation, and can do the job, so long as the currents
involved are not too strong compared with a scale that will
be estimated in Section \ref{Annex3}.

It is to be commented that the evident incompatibility of 
cylindrical symmetry with a non degenerate mapping into a 
curved (spherical or more general) target space means that for a
string loop retaining several non-commuting currents
it will be impossible (as in the simple Goto-Nambu case)
to attain a stationary circular vorton type equilibrium state:
such a loop will be condemned to go on oscillating until
completion of the dissipation of all currents except those of
an Abelian subgroup.

\section{Harmonious field models}
\label{Sec1}

According to the definition of the preceeding article \cite{III},
a multiscalar field $\ov\Phir$ with local components $\chir^{_\Ar}$
in a $q$-dimensional target space over a brane worldsheet $\calS$ 
of dimension $d$ (for $d\leq n$, where $n$ is the dimension of the 
background spacetime)will be of {\it harmonious} type if it is 
governed by a scalar Lagrangian $\ov\Lr$ that is specified by some 
equation of state as a diffeomorphically invariant function  just 
of the relevant target space metric $\hat\grd_{_{\Ar\Br}}$ -- which 
is supposed to have been prescribed in advance -- and of the 
corresponding horizontal projection $\hat\wrd{^{_{\Ar\Br}}}$ of the 
inverse $\ov\ggn{^{\,ij}}$ of the underlying spacetime metric 
$\ov\ggn_{ij}$ on the worldsheet. It is to be understood that the 
horizontality of the projection is specified with respect to a some 
gauge form $\ov\Aru_i$ with vectorial components $\ov\Aru_i{^{_\Ar}}$ 
on the target space. We thereby obtain a prescription of the form
{\be \hat\wrd{^{_{\Ar\Br}}}=\ov\ggn{^{\,ij}}\, \ov\Phir^{_\Ar}_{\ \wru\,i}\,
 \ov\Phir^{_\Br}_{\ \wru\,j}\, ,\label{1_1}\fe}
in which the relevant projector will be specified 
in terms of the field gradient components $\chir^{_\Ar}_{\, ,\, i}$ by
{\be \ov\Phir^{_\Ar}_{\ \wru\,i}=\chir^{_\Ar}_{\, ,\, i}
+\ov\Aru{_i^{\ _\Ar}}\, .\label{1_2}\fe}

In terms of the symmetric target space tensor $\kappar_{_{\Ar\Br}}$
given by the definition
{\be\kappar_{_{\Ar\Br}}=  -2\,\frac{\partial\ov\Lr}
{\partial\hat\wrd{^{_{\Ar\Br}}}}\label{1_3}\fe}
and therefor such that
{\be \kappar_{_\Cr}^{\,\ _\Ar}\,\hat\wrd^{_{\Br\Cr}}=
\kappar_{_\Cr}^{\,\ _\Br}\,\hat\wrd{^{_{\Ar\Cr}}} =
-2\,\frac{\partial\ov\Lr}{\partial\hat\grd_{_{\Ar\Br}}}
 ,\label{1_4b}\fe}
it was shown in the preceeding article \cite{III} that, for a system of 
this harmonious type, the intrinsic dynamical equations will
be expressible in terms of a set of surface currents with components 
given by
{\be \ov\Jr_{\!_\Ar}{^i}=\ov\ggn{^{\,ij}}\,\kappar_{_{\Ar\Br}} 
\,\ov\Phir^{_\Br}_{\ \wru\,j}\label{1_4}\, .\fe}
as the  pseudo-conservation laws
{\be \ov\Dru_{ i}\,\ov\Jr_{\!_\Ar}{^i}= 0\, ,\label{1_5}\fe}
in which $\ov\Dru_{i}$ is a bitensorially covariant 
differentiation operator of the not so simple kind introduced
 in earlier work \cite{I,II}, and which will give
rise to genuine current conservation laws \cite{III}
only when suitable internal symmetry conditions are satisfied.

One of the questions that arises in the study of any system of 
differential equations of motion is the orientation of the 
characteristic surfaces, with normal direction $\lambda_i$ say,
along which infinitesimal discontinuities can be propagated. 
For the transverse ``wiggle'' perturbation modes of the extrinsic 
evolution of the supporting worldsheet, it is easily  shown 
\cite{C95} that -- regardless of the internal dynamics -- the 
relevant characteristic equation will always have the simple form
{\be \ov\Tr{^{ij}}\,\lambda_i\,\lambda_j=0\, ,\label{1_6}\fe}
where $\ov\Tr{^{ij}}$ are the components of the surface
stress energy tensor, which for a harmonious system of the kind under
consideration here will be given \cite{III} by
{\be\ov\Tr{^{ij}}=\kappar_{_{\Ar\Br}}\,\ov\Phir{^{_\Ar\,\wru\,i}}
\,\ov\Phir{^{_\Br\,\wru\,j}}+\ov\Lr\,\ov\ggn{^{\,ij}}\, .\label{1_7}\fe}
However for the internal perturbation modes, wihin the worldsheet,
the form of the characteristic equation
will depend on the specific details of the system. In particular
for the accoustic type modes of the harmonious system (\ref{1_5})
the characteristic equation will in general be rather complicated,
reducing to a simple quadratic form like that of (\ref{1_6})
only under special conditions, such as those of the
simply harmonious and perfectly harmonious categories that will be
presented  in the following sections.

\section{Harmoniously elastic models}
\label{Sec2}

The category of harmonious models includes models of the perfect 
solid type \cite{CQ72} which belong to the extensive 
{\it harmoniously elastic} subcategory that is characterised
by the condition that the component matrix $\hat\wrd^{_{\Ar\Br}}$
should have a well defined inverse $\hat\gammar_{_{\Ar\Br}}
=\hat\wrd^{-1}_{\,_{\Ar\Br}}$, which, if it exists, will be
interpretable as the tensorially well behaved metric that is 
locally induced on the target space by the section $\Phir$ 
according to the specification
{\be \hat \gammar_{_{\Ar\Cr}}\hat \wrd^{_{\Cr\Br}}=
\delta_{\!_\Ar}^{\, _\Br}\, ,\label{1_9}\fe} 
This is something that will be possible only if the target-space 
dimension $q$ does not exceed the dimension $d=p+1$ of the 
supporting base space, which if it is an embedded $p$-brane 
worldsheet can itself not exceed the dimension $n$ of the 
background spacetime: $q\leq p+1\leq n$. The dimensionally
maximal case $q=p+1$ includes various models of the recently 
investigated kind \cite{C07} referred to a hyperelastic, while 
perfect solids\cite{CQ72} and so, in particular, ordinary fluid 
models of the barotropic type, are included in the case $q=p$.

Harmoniously elastic models are perfectly elastic in the usual 
sense, meaning \cite{C73a,C83} that they are governed by a Lagrangian that 
is determined by some prescribed intrinsic structure as a function 
just of the induced metric $\hat\gammar_{_{\Ar\Br}}$ on the relevant 
target space, but they are not elastic models of the most general kind: 
for non-harmoniously elastic models  the prescribed intrinsic 
structure can include various other vectorial or tensorial fields 
(for example to allow for the anisotropic grain in wood) as well as 
the prescribed metric $\hat\grd_{_{\Ar\Br}}$ which is all that is 
allowed in the harmonious case. In the usual approach 
\cite{CQ72,C73a,C83,CCC06} to the treatment of elastic solid models, the 
locally induced metric  $\hat\gammar_{_{\Ar\Br}}$ is what is used for 
lowering and raising of target space indices. It is therefor important 
to remember that in the dimensionally unrestricted approach 
\cite{I,II,III} followed here it is instead the globally prescribed 
target space metric $\hat\grd_{_{\Ar\Br}}$ that it is used for this 
purpose. 

It will be convenient for the following discussion  to introduce 
a new kind of elasticity tensor that is defined, for any 
harmonious model, by
{\be \Ematr_{_{\Ar\Br\Cr\Dr}}= \Ematr_{_{(\Cr\Dr)(\Ar\Br)}}
= 2\frac{\partial\kappar_{_{\Ar\Br}}}{\partial \hat \wrd{^{_{\Cr\Dr}}}}\
\, .\label{1_10}\fe}
In terms of this quantity the ordinary elasticity tensor of the usual 
treatment \cite{CQ72,C73a,C83,CCC06} will be given by the expression
{\be \Er_{_{\Ar\Br\Cr\Dr}}= \Ematr_{_{\Ar\Br\Cr\Dr}}+ 2\kappar_{_{\Ar(\Cr}}
\gammar_{_{\Dr)\Br}}-\kappar_{_{\Ar\Br}}\gammar_{_{\Cr\Dr}}
+2\gammar_{_{\Ar(\Cr}}\Prd_{_{\Dr)\Br}}-\gammar_{_{\Ar\Br}}
\Prd_{_{\Cr\Dr}}\, ,\label{1_11}\fe}
in which the pressure tensor is
given in terms of the energy density $\rhor=-\Lr$ by
{\be \Prd_{_{\Ar\Br}}=\kappar_{_{\Ar\Br}}-\rhor\,\gammar_{_{\Ar\Br}}
\, .\label{1_12}\fe}
A simple example \cite{III} is that of the baby Skyrme model
\cite{HK08} for which the target space is a 2-sphere on which 
$ \kappar_{_{\Ar\Br}}=\kapparo_\star\grd_{_{\Ar\Br}}
+\alpharo_\star(\hat\wrd\,
\grd_{_{\Ar\Br}}-\hat\wrd_{_{\Ar\Br}})$,
where $\kapparo_\star$ and $\alpharo_\star$ are constants,
so that one obtains $\Ematr_{_{\Ar\Br\Cr\Dr}}=2\alpharo_\star
(\grd_{_{\Ar\Br}}\grd_{_{\Cr\Dr}}-\grd_{_{\Ar(\Cr}}\grd_{_{\Dr)\Br}})$.

\section{The sound cone in simply harmonious models}
\label{Sec3}

Instead of working through the details of a complete perturbation 
analysis, the characteristic equation governing the propagation
of infinitesimal discontinuities is obtainable  efficiently
 by adaption, from  the rather similar case 
of an elastic solid \cite{CCC06,C73}, of a method due originally 
to Hadamard, of which the simplest illustration is provided by the 
Dalembertian wave equation for a scalar $\varphir$ say, namely 
$\ov\nablag_{\! j}\,\varphir{^j}=0$ where $\varphir{^j} =\ov
\ggn{^{\,ji}}\ov\nablag_{\! i}\varphir$. The idea of the Hadamard
method is to use the fact that the discontinuity of the gradient 
of a continuous quantity will be aligned with the normal covector
$\lambda_i$ of the discontinuity surface. Applying this to the
components $\varphir{^j}$, one sees that the discontinuity of
their gradients will be given in terms of corresponding
discontinuity amplitude components $\widetilde\varphir{^j}$
by an expression of the form $[\nablag_{\! i}\,\varphir{^j}]
=\lambda_i \widetilde\varphir{^j}$. Moreover, taking the 
discontinuity of the integrability condition
$\ov\nablag_{\! [i}\varphir_{j]}=0$, one sees that the discontinuity
amplitude will have to satisfy $\lambda_{[i}\widetilde\varphir_{j]}
=0$, and hence that it will be given in terms of some scalar 
amplitude $\widetilde\varphir$ by $\widetilde\varphir{_i}=
\widetilde\varphir\lambda_i$. Thus finally, taking the 
discontinuity of the the Dalembert equation itself, one obtains 
the well known light-cone tangency condition 
{\be \ov\ggn{^{\,ij}}\,\lambda_i\, \lambda_j =0\, .\label{2_0}\fe}

Applying the same line of reasoning to the multiscalar field
$\ov\Phir$, one sees that the discontinity of the gradient 
of its covariant derivative (\ref{1_2}) will be given in terms 
of some set of amplitude components $\widetilde\chir{^{_\Ar}}$ by
{\be [\ov\nablag_{\! i}\,\ov\Phir^{_\Ar}_{\ \wru\,j}]=
\lambda_i\, \lambda_j \,\widetilde\chir{^{_\Ar}}\, .\label{2_1}\fe}
It therefore follows from the definition (\ref{1_1})
of the horizontally induced metric $\ov \wrd{^{_{\Ar\Br}}}$
that discontinuity of its gradient will be given by
{\be [\ov\nablag_{\! i}\, \hat \wrd{^{_{\Ar\Br}}}]=2
\lambda_i\, \lambda_j \,\widetilde\chir{^{_{(\Ar}}}
\ov\Phir{^{_{\Br)}\,\wru\,j}}\, .\label{2_2}\fe}
These quantities are needed for the evaluation of the discontinuity
of the gradient of the current  (\ref{1_4}), which will be given by
the formula
{\be[\ov\Dru_{ i}\,\ov\Jr_{\!_\Ar}{^j}]=\kappar_{_{\Ar\Br}}\,
\ov\ggn{^{\,jk}}  [\ov\nablag_{\! i}\,\ov\Phir^{_\Br}_{\ \wru\,k}]+
\frac{\partial\kappar_{_{\Ar\Br}}}{\partial \hat \wrd{^{_{\Cr\Dr}}}}\,
\ov\Phir{^{_\Br\,\wru\,j}}\, [\ov\nablag_{\! i}\, \hat 
\wrd{^{_{\Cr\Dr}}}]\, .\label{2_3}\fe}
in which the distinction between $\ov\Dru_{ i}$  and 
$\ov\nablag_{\! i}$ disappears, as the relevent \cite{III} affine 
and gauge connection terms (but not their derivatives) will be 
continuous. The discontinuity of the set of pseudo-conservation 
equations (\ref{1_5}) thereby provides the required characteristic 
equation in the form
{\be \lambda_i\, \lambda_j  {\cal Q}^{ij}_{\ _{\Ar\Br}}
\widetilde\chir{^{_\Br}}=0 \, ,\label{2_4}\fe}
in which, using the notation (\ref{1_10}), we shall have
{\be  {\cal Q}^{ij}_{\ _{\Ar\Br}}=\ov\ggn{^{\,ij}}
\kappar_{_{\Ar\Br}}+\Ematr_{_{\Ar\Dr\Br\Cr}}\,\ov
\Phir{^{_\Cr\,\wru\,i}}\,\ov\Phir{^{_\Dr\,\wru\,j}} 
\, \label{2_5}.\fe}
It is to be remarked that this formula is simpler than the 
corresponding expression using the usual 
elasticity tensor (\ref{1_11}) of the traditional approach 
\cite{CQ72,C73a,C83,CCC06}.

The eigenvalue equation ensuing from (\ref{2_4}) may take a rather 
complicated quartic or higher polynomial form when the target space 
dimension is two or more, with a generic equation of state involving 
dependence not just on the  trace
{\be \hat \wrd=\hat \wrd_{_\Ar}{^{_\Ar}}=\hat \grd_{_{\Ar\Br}}
\hat \wrd^{_{\Ar\Br}}\, ,\label{2_6}\fe}
but also on higher order invariants starting
with $\hat \wrd_{_\Ar}{^{_\Br}}\hat \wrd_{_\Br}{^{_\Ar}}$.
However it will conveniently separate into merely quadratic
subsystems in what will be referred
to as the {\it simply harmonious} case, namely that for which
the Lagrangian depends {\it only} on the trace invariant $\hat\wrd$,
so that one obtains
{\be\kappar_{_{\Ar\Br}}=\kappar\,\hat\grd_{_{\Ar\Br}}
\, ,\label{2_7}\fe}
with
{\be \kappar=-2 \,\frac{{\rm d}\ov\Lr}{{\rm d}\hat\wrd} 
\, .\label{2_8}\fe}
In this simply harmonious case one obtains
{\be \Ematr_{_{\Ar\Br\Cr\Dr}}=2\frac{{\rm d} \kappar} {{\rm d} 
\hat\wrd}\, \hat\grd_{_{\Ar\Br}}\,\hat\grd_{_{\Cr\Dr}}
\, ,\label{2_9}\fe}
which  gives a characteristic equation of the form
{\be\lambda_i\lambda^i\,\widetilde\chir{^{_\Ar}}+
\frac{2}{\kappar}\frac{{\rm d}\kappar}{{\rm d}\hat\wrd}\,
\lambda^i\ov\Phir{^{_\Ar}_{\ \wru\,i}\,\lambda^j} 
\ov\Phir{^{_\Br}_{\ \wru\,j}}\,\widetilde
\chir_{_\Br}=0\, .\label{2_10}\fe}

It is evident that this will be trivially satisfied by  a set of 
shake modes, propagating at the speed of light, with polarisation
$\widetilde\chir{^{_\Ar}}$ that is transverse to the current across 
the discontinuity, in the sense that 
{\be\widetilde\chir{^{_\Ar}}
\ov\Jr_{\!_\Ar}{^i}\lambda_i=0\, ,\label{2_11}\fe}
since for such a mode -- regardless of the particular linear or 
non-linear functional form of the equation of state -- the 
characteristic equation will evidently reduce just to the same 
nullity condition as in the ordinary Dalembertian case (\ref{2_0}), 
namely
{\be \lambda_i\lambda^i=0\, .\label{2_12}\fe}
There will also be a less trivial set of sound type modes with 
polarisation that is longitudinal in the sense of being aligned 
with the current across the discontinuity, so  that  for such a mode 
the discontinuity amplitude vector $\widetilde\chir{^{_\Ar}}$ will be
given (modulo a multiplicative factor that can be absorbed into the 
normalisation of the characteristic covector $\lambda_i$)  by
the prescription
{\be\widetilde\chir{^{_\Ar}}=\ov\Phir{^{_\Ar}_{\ \wru\,i}}\,
\lambda^i\, .\label{2_13}\fe}
This reduces the characteristic equation to a simple quadratic
form -- specifying what is describable as a sound cone -- that will
be given by
{\be \left(\ov \ggn_{ij}+\frac{2}{\kappar}
\frac{{\rm d}\kappar}{{\rm d}\hat\wrd}\,\ov \wgn_{ij}\right)
\lambda^i\lambda^j=0 \label{2_14}\fe}
using the notation
{\be \ov \wgn_{ij}=\hat\grd_{_{\Ar\Br}}
\ov\Phir^{_\Ar}_{\ \wru\,i}\ov\Phir^{_\Br}_{\ \wru\,j}
\,\label{2_15}\fe} 
for the gauge covariant pullback of the target space metric.

It will be shown below how non-trivially non-linear models of this
simply harmonious type arise naturally  in the treatment of string 
defects of multiscalar field theories of the common kinetically linear 
kind. However before doing that it is will be instructive, for the 
sake of comparison, to describe another noteworthy subcategory, namely 
that of {\it perfectly harmonious} models for which the characteristic 
equation will be similarly simplifiable

\section{The sound cone in perfectly harmonious models}
\label{Sec4}

The subcategory of what are describable as {\it perfectly harmonious} 
models is physically important because it includes the generic 
(not necessarily irrotational) case of an ordinary barotropic 
perfect fluid. The perfectly harmonious subcategory is defined by the 
requirement that the dependence of the Lagrangian
on the target space tensor $\hat \wrd^{_{\Ar\Br}}$ should
again involve only a single scalar invariant, but with the latter
now chosen to be the determinant ${\rm det}\{\hat\wrd\}$
of the matrix with components
{\be \hat\wrd_{_\Ar}{^{_\Br}}=\hat \grd_{_{\Ar\Cr}}
\hat \wrd^{_{\Cr\Br}} \, ,\fe}
which will be admissible so long as the target space dimension $q$
does not exceed the base space-time dimension $d=p+1$ 
(whereas for $q>p+1$ this determinant would vanish identically).
Ordinary perfect fluids are of the particular kind for which the 
target space dimension is the same, $q=p$, as the space (as distinct 
from space-time) dimension of the supporting base, which is 3 in 
the usual terrestrial and astrophysical applications, but might be 
higher for exotic cosmological theories in which the space-time 
dimension is not 4 but 5 or more.

In terms of the tensorial inverse matrix $\hat\gammar_{_{\Ar\Br}}
=\hat\wrd^{-1}_{\,_{\Ar\Br}}$ of $\hat\wrd^{_{\Ar\Br}}$ (which is
interpretable as the metric locally induced on the target space by 
the section $\Phir$) as defined by (\ref{1_9}), the stipulation 
that $\Lr$ should depend only on ${\rm det}\{\hat\wrd\}$  leads to 
the expression 
{\be \kappar_{_{\Ar\Br}}=\hrd\,\hat\gammar_{_{\Ar\Br}}\fe}
with
{\be\hrd=
-2\, \frac{{\rm det}\{\hat\wrd\}\,{\rm d}\Lr}
{\, {\rm d}({\rm det}\{\hat\wrd\})}\, .\fe}
This quantity $\hrd$ will be interpretable simply as the enthalpy
density in the case of an ordinary perfect fluid, for which the 
pressure tensor (\ref{1_12}) takes the form
$ \Prd_{_{\Ar\Br}}=\Prd\,\hat\gammar_{_{\Ar\Br}}$,
in which the pressure is given in terms of the energy density  
$\rhor=-\Lr$,  and the enthalpy density $\hrd$ by the well known 
formula  $\Prd =\hrd-\rhor$.

The ensuing formula
 {\be \frac{\partial\kappar_{_{\Ar\Br}}}
{\partial\hat\wrd{^{_{\Cr\Dr}}}}=\frac{{\rm det}\{\hat\wrd\}
\,{\rm d}\hrd}{\, {\rm d}({\rm det}\{\hat\wrd\})}\,
\hat\gammar_{_{\Ar\Br}}\hat\gammar_{_{\Cr\Dr}}-\hrd\,
\hat\gammar_{_{\Ar(\Cr}}\hat\gammar_{_{\Dr)\Br}}\, ,\fe}
can be used to reduce the characteristic matrix (\ref{2_5})
to the form 
{\be{\cal Q}^{ij}_{\ _{\Ar\Br}}=\hat\gammar_{_{\Ar\Br}}\left(\hrd\,
\ov\ggn{^{\,ij}}-\kappar_{_{\Cr\Dr}}\ov\Phir{^{_\Cr\,\wru\,i}}
\,\ov\Phir{^{_\Dr\,\wru\,j}}\right)+ \Big(\! 2\frac{{\rm det}
\{\hat\wrd\}\,{\rm d}\hrd}{\, {\rm d}({\rm det}\{\hat\wrd\})}
\, -\hrd\!\Big)\hat\gammar_{_{\Ar\Cr}}\ov\Phir{^{_\Cr\,\wru\,i}}
\hat\gammar_{_{\Br\Dr}}\ov\Phir{^{_\Dr\,\wru\,j}}\, .\fe}

As before, this will be  satisfied by trivial shake modes, with 
polarisation $\widetilde\chir{^{_\Ar}}$ that is transverse to the 
current across the discontinuity in the sense specified by 
(\ref{2_11}), since for such a mode -- regardless of the particular 
linear or non-linear functional form of the equation of state -- 
the characteristic equation will  reduce to the quadratic form
{\be \left((\hrd+\ov\Lr)\ov\ggn{^{\,ij}}-\ov\Tr{^{ij}}\right)
\lambda_i\lambda_j=0\, ,\fe}
with $\ov\Tr{^{ij}}$ as given by (\ref{1_7}). 

In the ordinary perfect fluid case this will take 
the degenerate form $(\bar\ugn{^i}\,\lambda_i)^2=0$, where 
$\bar\ugn{^i}$ is the timelike (and physically well defined) 
unit fluid flow tangent vector that is characterised 
by the  condition $\ov\Jr{^{_\Ar}_{\  i}}\,\bar\ugn{^i}=0$, 
meaning orthogonality to all the (separately unphysical, since 
target coordinate dependent) currents, and in terms of which the 
stress-energy tensor will take the familiar form  $\ov\Tr{^{ij}}=
\hrd\, \bar\ugn{^i}\,\bar\ugn{^j}+\Prd\,\ov\ggn{^{\,ij}}$.

As before, there will also be a  set of non-trivial sound type modes 
with polarisation that is longitudinal in the sense specified by
(\ref{2_13}), for which the characteristic equation will reduce to 
the quadratic form
{\be \left(\hrd\,\ov\ggn{^{\,ij}}+2\Big( \frac{{\rm det}
\{\hat\wrd\}\,{\rm d}\hrd}{\,\hrd\, {\rm d}({\rm det}
\{\hat\wrd\})}\, -1\!\Big)(\ov\Tr{^{ij}}-\Lr\,\ov\ggn{^{\,ij}})
\right)\lambda_i\lambda_j=0\, ,\fe}
which is what characterises the ordinary sound cone in the
familiar perfect fluid case.

\section{Extended Witten models}
\label{Annex1}

The physical relevance of the perfectly harmonious category 
presented in the immediately preceeding section is obvious, at 
least in the case of the ordinary elastic solid and fluid 
applications for which the target space dimension is $q=3$.
However for the study of the simply harmonious category, as 
presented in the section before that, some physical motivation 
needs to be provided. In the case for which the target space is 
one dimensional (and for which simply harmonious means the same 
thing as perfectly harmonious) such a justification was provided 
by the demonstration \cite{CP95,HC08} that such models are what 
is appropriate for the macroscopic description of string defects 
in simple kinetically linear field models of a subcategory 
proposed by Witten\cite{Witten85}. The purpose of the present 
section is to present a straightforward extension of Witten's 
subcategory that can form string defects which will be shown in 
the following section to be macroscopically describable by
simply harmonious models of a less trivial kind, with two -- or 
higher--  dimensional target spaces that are curved.

Within the category characterised by a Lagrangian of the 
forced-harmonic type \cite{III}, the original Witten subcategory 
and the extensions considered here are characterised by two 
essential properties of which the first is  that of having a 
$(3+q)$-dimensional target space $\calXr$ of the ordinary flat 
kind, so that the symmetry group of the kinetic part $\Lr_{\bf kin}$ 
of the Lagrangian is 0$\{3+q\}$. The second property is that the 
target space is however endowed with a potential function 
$\hat\calVr$ depending on just two scalar combinations, namely a 
squared ``Higgs amplitude'' $\Psir^2$ and  a squared ``carrier 
amplitude''$\Sigmar^2$. These are obtained by decomposing the 
target space as a direct product of a 2-dimensional ``Higgs field'' 
space and a $(q+1)$-dimensional ``carrier field'' space, so that 
the symmetry group of the whole Lagrangian, 
{\be\Lr=\Lr_{\bf kin} -\hat\calVr\, ,\label{annex0}\fe}
will be generically broken down to the direct product,  
0$\{2\}\times$0$\{q+1\}$, of a Higgs field symmetry group having 
the form 0$\{2\}$ with a ``carrier'' symmetry group having 
the form 0$\{q+1\}$. The idea is that the potential should be 
such that the  0$\{2\}$ symmetry of the Higgs part is spontaneously 
broken, so that the vacuum will admit the occurrence of string type 
topological defects (which will be ``local'' if the  symmetry 
algebra of the ``Higgs field'' part is ``gauged'') containing fields 
whose internal symmetries are just those of the carrier group  
0$\{q+1\}$.

The ``Higgs field'' space can be taken to have flat coordinates 
$\Psir^{_1}$ and $\Psir^{_2}$ say, which can be conveniently 
thought of as the real and imaginary parts of a complex field
{\be \Psir^{_1}+i\Psir^{_2}=\Psir \,{\rm e}^{i\psir} 
\, ,\label{annex_1}\fe} 
in terms of which the  2-dimensional Higgs field  part 
of the target space metric will be 
{\be {\rm d}\hat\srd_{\bf hig}^2={\rm d}\Psir^{_1\,2}+{\rm d}
\Psir^{_2\,2}={\rm d}\Psir^2+\Psir^2 {\rm d}\psir^2
\, ,\label{annex5}\fe}
and corresponding Higgs field amplitude will be given by
{\be  \Psir^2=\Psir^{_1\,2}+\Psir^{_2\,2}\, .\label{annex_2}\fe}
Similarly the ``carrier field'' space can be taken to have flat 
target space coordinates, $\Sigmar^\ard$ say, $\ard=1, ..., q+1$  
in terms of which the carrier metric contribution will be 
{\be {\rm d}\hat\srd_{\bf car}^2=\delta_{\ard\brd}\, {\rm d}
\Sigmar^\ard\,{\rm d}\Sigmar^\brd \, ,\label{annex6}\fe}
where $\delta_{\ard\brd}$ is the unit matrix,
and the carrier amplitude itself will be given by 
{\be \Sigmar^2=\delta_{\ard\brd}\, \Sigmar^\ard\Sigmar^\brd
\, .\label{annex_3}\fe}
These contributions combine to give the complete metric on the
 flat target space $\calXr$ as
 ${\rm d}\hat\srd_{\bf hig}^2+{\rm d}\hat\srd_{\bf car}^2$, which 
means that the kinetic part of the Lagrangian will take the form
{\be \Lr_{\bf kin}=\Lr_{\bf hig}+\Lr_{\bf car}\, ,
\label{annex_3a} \fe} 
with
{\be \Lr_{\bf hig}= -\frac{_1}{^2}\Big( (\Dru_\mu\Psir^{_1})
\Dru^\mu\Psir^{_1}+(\Dru_\mu\Psir^{_2})
\Dru^\mu\Psir^{_2}\Big)-\frac{
\Fgoru_{\mu\nu}\Fgoru^{\mu\nu}}{16\pi\egothrd^2}
\, ,\label{annex_4}\fe}
and
{\be \Lr_{\bf car}=-\frac{_1}{^2}\delta_{\ard\brd}(\nablab_{\!\mu} 
\Sigmar^\ard)\nablab^\mu\Sigmar^\brd\, ,\label{annex_4a}\fe}
where the internal gauge coupling of the Higgs field 
part has been  incorporated by the use of the covariant 
differentiation operation that is given by 
$$\Dru^\mu\Psir^{_1} =\nablab_{\!\mu}\Psir^{_1} -\Agoru_\mu\Psir^{_2}
\, ,\hskip 1 cm  \Dru^\mu\Psir^{_2}=\nablab_{\!\mu}\Psir^{_2} +
\Agoru_\mu\Psir^{_1}\, ,$$ 
where $\Agoru_\mu$ is a U\{1\} gauge form with curvature 
$\Fgoru_{\mu\nu}=2\nablab_{\![\mu}\Agoru_{\nu]}$, for which Gothic 
letters have been used to indicate that, although mathematically
analagous, this internal gauge field is not meant to be physically 
interpretable as the ordinary electromagnetic field. 
For a small but non-zero value of the coupling constant $\egothrd$
this gauge field enables the vortex defects of the model to be 
locally confined -- without the logarithmic energy divergence for 
which a long range ``infra red''  cut off would otherwise be needed.

In his original formulation \cite{Witten85} Witten made the further 
postulate that, as well as this internal gauge coupling of the Higgs
field, there would also be an external gauge coupling of
the``carrier field'' part  to an analogous  
U\{1\} gauge form $\Aru_{\nu}$, with curvature 
$\Fru_{\mu\nu}=2\nablab_{\![\mu}\Aru_{\nu]}$, that was meant to be 
interpreted as that of an ordinary electromagnetic field, with 
its own extra Lagrangian contribution $\Fru_{\mu\nu}\Fru^{\nu\mu}
/16\pi\erd^2$. However -- unless the corresponding coupling 
constant, $\erd$ say, is set to zero -- such a coupling engenders 
technical trouble by reintroducing the logarithmic ``infra-red'' 
divergence that had been removed by the other gauge coupling.

The present treatment will be based on the supposition that gauge 
self-coupling of the ``carrier field'' part is weak enough to be 
neglected, so that the divergence problem is avoided, but this 
does not exclude allowance for passive coupling to an external 
background of electromagnetic or other conceivable radiation. 
It will nevertheless be supposed that such radiation is 
sufficiently weak to allow the gauge to be chosen so that the 
corresponding gauge form (namely $\Aru_{\nu}$ in the 
electromagnetic case) to be taken to be zero in a neigbourhood 
that is large compared with the internal dimensions of the defect, 
so that within this neighbourhood there will be no further loss of 
generality in taking the kinetic part of the Lagrangian to have 
the simple form (\ref{annex_4}).


The original Witten model was characterised by $q=1$, so that the 
carrier target space coordinates could be considered as components
of a  complex field $\Sigmar^{_1}+i\Sigmar^{_2}=\Sigmar\,{\rm e}^{i\chir}$. 
What I refer to as the minimally extended  Witten model is 
characterised by $q=2$, so that the carrier symmetry group will have 
the non-Abelian form   0$\{3\}$, instead of the Abelian form  0$\{2\}$ 
that it had in the original Witten model. 

In a more elaborate -- non-minimal -- extension proposed for 
consideration by Lilley {\it et al.} (private communication 
\cite{LMMP09}) the carrier space dimension is taken to be given by 
$q+1=4$. Instead of retaining the its full symmetry group 0$\{4\}$, 
these authors took the carrier space to be endowed with a complex 
structure by grouping its 
coordinates into a pair of complex fields $\Sigmar^{_1}+i\Sigmar^{_2}$ 
and $\Sigmar^{_3}+i\Sigmar^{_4}$, so that the carrier symmetry group 
is reduced to the form U$\{2\}$. The latter has the structure of a 
direct product of an SU$\{2\}$ group (which Lilley {\it et al.} took 
to be ``gauged'') with a U$\{1\}$ group  (which they left as merely 
``global'', but which could just as well be taken to be coupled
to ordinary electromagnetism).

In all these cases the flat metric (\ref{annex6}) of the $(q+1)$-dimensional 
carrier part can be rewritten as
{\be {\rm d}\hat\srd_{\bf car}^2={\rm d}\Sigmar^2+ 
\Sigmar^2 {\rm d}\hat\Omegar^2 \fe}
where   ${\rm d}\hat\Omegar^2$ is the metric on the relevant symmetry-orbit 
space, $\ov\calXr$, which will be the unit $q$-sphere
as given in terms of some system of coordinates $\chir^{\!_\Ar}$
by an expression of the form
{\be {\rm d}\hat\Omegar^2=\hat\grd_{_{\Ar\Br}}\, {\rm d}\chir^{\!_\Ar}\,
{\rm d}\chir^{\!_\Br}\, .\label{annex7}\fe}
More particularly, for the minimally extended model characterised
by $q=2$, the standard choice 
$\chir^{_1}=\hat\thetar$ and $\chir^{_2}=\hat\varphir$ 
will be obtained by setting $\Sigmar^{_1}=\Sigmar\, {\rm sin}\, \hat\thetar
\,{\rm cos}\,\hat\varphir$, $\Sigmar^{_2}=\Sigmar\, {\rm sin}\, \hat\thetar
\,{\rm sin}\,\hat\varphir$, and   
$\Sigmar^{_3}=\Sigmar\, {\rm cos}\, \hat\thetar$.
In terms of these, the spherical metric components will be given  by a 
prescription of the familiar form
$\hat \grd_{_{11}}=1$,  $\hat \grd_{_{12}}=0$, 
$\hat \grd_{_{22}}= {\rm sin}^2\, \hat\thetar$. 

The basic idea behind cosmic string theory, as developed at first most 
notably by Kibble \cite{Kibble76,KLY97} , was that short, effectively 
straight string segments in a locally uniform background neighbourhood 
could be approximated by Nielsen Olesen type vortex solutions of the 
underlying field model.The presence of longitudinal currents in the 
vortex was excluded by the rather strong kind of cylindindrical 
symmetry postulated by an ansatz of the Nielsen Olsen type, but was 
admitted by a weaker kind of cylindrical symmetry ansatz that was 
subsequently introduced by Witten \cite{Witten85}. As shown by the 
recent work of Lilley {\it et al.} (private communication 
\cite{LMMP09}) even the weaker kind of cylindical 
symmetry ansatz proposed by Witten is incompatible with the simultaneous 
presence of several non-commuting longitudinal currents, whose treatment
will therefor require an ansatz of an even weaker kind that will be
introduced in the next section,
whereby it is required  that the cylindrical symmetry should hold only 
approximately in the relevant locally uniform background neigbourhood.

Provided the external gauge coupling is sufficiently weak for its 
self coupling to be neglected, it will be possible to choose the gauge 
so that the corresponding (electromagnetic) guage form $\Aru_{\nu}$
vanishes in the neighbourhood under consideration. The field equations 
for the remaining internal guage form $\Aru_{\nu}$ and for the flat 
target space components will then be expressible in kinetically 
decoupled form as a first subsystem consisting of
{\be  \nabla_{\!\nu}\Fgoru^{\mu\nu}=4\pi\egothrd^2\,\big(
\Psir^{_2}\Dru^\mu\Psir^{_1}-\Psir^{_1}\Dru^\mu\Psir^{_2}\big)
\, ,\label{annex11a}\fe}
and
{\be \Dru_{\mu}\Dru^\mu\,\Psir^{_1}=2\,\frac{\partial\hat\calVr}
{\partial(\Psir^2)}\ \Psir^{_1}\, ,\hskip 1 cm
 \Dru_{\mu}\Dru^\mu\,\Psir^{_2}=2\,\frac{\partial\hat\calVr}
{\partial(\Psir^2)}\ \Psir^{_2}\, ,\label{annex11}\fe}
for the gauge form and the Higgs field, and a second subsystem 
given simply by
{\be \nabla_{\!\mu}\nabla^\mu\,\Sigmar^{\ard}=2\,\frac{\partial
\hat\calVr}{\partial(\Sigmar^2)}\ \Sigmar^{\ard}
\, ,\label{annex12}\fe}
(in which $\ard=1, ..., q+1$) for the carrier field.

\section{Witten's weak symmetry ansatz}
\label{WittenAnsatz}

Following  Witten \cite{Witten85}, attention  will now
be restricted to configurations in which the Higgs
subsystem is subject to the same symmetry conditions as in
a simple Nielsen-Olesen type vortex (for which the carrier 
subsystem is absent). This means that
in a flat spacetime, with respect to a cylindrical coordinates
for which the metric is
{\be {\rm d}\sbe^2={\rm d}\varrhob^2+\varrhob^2 {\rm d}\phi^2
+{\rm d}z^2-{\rm d}t^2\, ,\label{annex13}\fe}
the Higgs field and its gauge form are postulated to be
longitudinally symmetric in the strong sense, to the effect that 
$\Psir^{_1}$ and $\Psir^{_1}$ are independent of $z$ and $t$, 
but they are required to be axially symmetric only in the weak 
(albeit strict \cite{III}) sense, meaning modulo an action 
of the primary symmetry group O$\{2\}$, so that the phase $\psir$ 
in (\ref{annex_1}) is allowed to have an angle dependence of the form
{\be \psir=\nrd\,\phi\, ,\label{annex14}\fe}
where $\nrd$ is a fixed integer winding number, while the
amplitude $\Psir$ can depend only on $\varrhob$. The corresponding 
ansatz for the internal gauge field is that it should have the form
{\be \Agoru_{\mu}\, {\rm d}x^\mu=\Agoru\, {\rm d}\phi
\, ,\label{annex14b}\fe}
in which the quantity $\Agoru$ is also a function only of  $\varrhob$.


Still following  Witten \cite{Witten85}, it will be postulated
that the carrier subsystem is axisymmetric in the strong sense 
-- meaning that the field components $\Sigmar^\ard$ are all 
independent of $\phi$ --so that using a prime for differentiation with 
respect to $z$ and a dot for differentiation with respect to $t$,
their dynamical equations (\ref{annex12}) will take the form
{\be \frac{1}{\varrhob}\frac{{\rm d}}{{\rm d}\varrhob}\left(
\varrhob \frac{{\rm d}\Sigmar^\ard}{{\rm d}\varrhob}\right)
=\ddot\Sigmar{^\ard}-\Sigmar^{\prime\prime \ard}
+2\,\frac{\partial\hat\calVr}
{\partial(\Sigmar^2)}\ \Sigmar^{\ard}\, ,\label{annex16}\fe}

An ansatz of the restrictive  Nielsen Olesen type postulated by 
Kibble \cite{Kibble76} would also require staticity and cylindrical 
symmetry in the strong sense, meaning 
$\dot\Sigmar{^\ard}=0$ and $\Sigmar^{\prime \ard}=0$, so that the 
components $\Sigmar^\ard$ should  depend only on $\varrhob$. The  
system of field equations would thereby be reduced from four to 
two dimensions, namely those of a flat cross section with fixed 
longitudinal coordinate values that can, without loss of generality, 
be taken to be $t=0$ and $z=0$. Macroscopic quantities such as 
the string energy per unit length will then be obtainable 
by integration over the cross section.

Witten's innovation \cite{Witten85} was to recognise that such a 
reduction to a two dimensional system on a flat cross section will 
still be obtainable from a less restrictive ansatz whereby the 
longitudinal symmetry of the carrier field is required to be only 
of the weak type. The Witten ansatz can be decomposed into
two successive conditions, of which the first is
 that the amplitude $\Sigmar$ can depend 
only on $\varrhob$, so that
{\be \dot\Sigmar=\Sigmar^\prime=0\, \label{annex17}\fe}
but that subject, of course, to the ensuing restaints, namely
{\be \delta_{\ard\brd}\, \Sigmar^\ard\dot\Sigmar{^\brd}=0\, ,
\hskip 1 cm \delta_{\ard\brd}\, \Sigmar^\ard\Sigmar^{\prime \brd}=0\
\, ,\label{annex18}\fe}
the longitudinal gradients $\dot\Sigmar{^\brd}$ and 
$\Sigmar^{\prime \brd}$ are allowed to have non vanishing values.
The second condition of the Witten ansatz is that these 
 gradient fields themselves should be 
longitudinally symmetric in the strong sense.
It is this second condition that will have to be relaxed in
the work that follows.



There will be no obstacle to the implementation of such a Witten 
type symmetry ansatz provided the relevant symmetry-orbit space
 $\ov\calXr$ say (meaning the quotient of the carrier field space by 
the action of its symmetry group) happens to be {\it flat} -- as in 
the single carrier component case originally considered by Witten, 
as well as in multicomponent Abelian cases that have been 
considered more recently \cite{LMP09}. In such cases a linear 
symmetry-preserving map fom the $z,t$ plane to $\ov\calXr$ will be 
available as a framework 
for parallel propagation
of the field  on the sample cross section at $t=0\, , \ z=0$,
so as to construct a solution that remains valid for all values 
of $t$ and $z$. Subject to the choice of a system of {\it flat} 
coordinates $\chir^{\!_\Ar}$ on the symmetry-orbit space $\ov\calXr$,
the Witten ansatz simply amounts to taking uniformly constant values
for  $\chir^{\!\prime_\Ar}$ and $\dot\chir{^{\!_\Ar}}$. The supplementary
requirement of invariance under the discrete symmetries of time
and parity reversal imply the further restriction that the initial
value of $\chir{^{\!_\Ar}}$ at $t=0\, ,\ z=0$  be uniform over
the cross section,  meaning independent not just 
of $\phi$ but also of $\varrhob$.

Such an ideal procedure will unfortunately  be available only for a 
restricted choice\cite{LMMP09} of the values of $\chir^{\!\prime_\Ar}$ 
and $\dot\chir{^{\!_\Ar}}$ if -- as in the cases we are concerned with 
here -- the relevant $q$-dimensional symmetry-orbit space $\ov\calXr$ 
is {\it curved} so that its symmetry algebra is non-Abelian, with the 
implication that Lie transport operations with respect to different 
generators will fail to be mutually consistent. In order to obtain 
an effectively 2 dimensional description on a sample cross section 
at $t=0$, $z=0$, in cases involving currents aligned with symmetry 
generators that do not commute, the Witten type weak symmetry ansatz
will neeed to be replaced by something less restrictive.
 
\section{The local geodicity ansatz}
\label{Annex2}

 For  the treatment of a generic current configuration, 
what I propose is something describable as the
{\it local geodicity} ansatz, whereby one is enabled to start  from 
{\it arbitrarily} chosen uniform values of, $\chir{^{\!_\Ar}}$ and its 
derivatives  $\chir^{\!\prime_\Ar}$ and  $\dot\chir{^{\!_\Ar}}$ on an 
initial cross section at  $t=0\, ,\ z=0$. This ansatz retains the 
first condition of the Witten ansatz, as embodied in (\ref{annex17}) 
and (\ref{annex18}), but instead of a further symmetry requirement
the second condition of the 
 local geodicity ansatz is that the value 
of $\chir{^{\!_\Ar}}$ throughout a finite spacetime neighbourhood 
of the cross section should be obtained by the standard process 
of geodesic extrapolation with respect to the metric 
$\hat\grd_{_{\Ar\Br}}$ on $\ov\calXr$. According to this prescription, 
a coordinate pair $\{z,t\}$ maps to a position specified by unit 
parameter value $\tau=1$ on the affinely parametrized geodesic 
$\chir{^{\!_\Ar}}\{\tau\}$ specified at $\tau=0$ by the tangent 
{\be \frac {{\rm d}\chir^{\!_\Ar}}{{\rm d}\tau}= t\,\dot
\chir{^{\!_\Ar}}+z\,\chir^{\!\prime_\Ar}\, .\label{annex19}\fe} 

So long as one is concerned with derivatives of at most second order, 
which is all that is needed for the field equations in question, the 
application of this ansatz is very easily achievable by choosing to 
work with local coordinates such that the relevant connection components 
vanish. In such a system the local geodicity ansatz simply means that 
all the second derivatives will also vanish: 
{\be \hat\Gammar_{\!_\Ar\ _\Cr}^{\ _\Br}=0\ \ \  \Rightarrow \ \ \
\ddot\chir{^{\!_\Ar}}=\dot\chir{^{\!\prime_\Ar}}
=\chir{^{\!\prime\prime_\Ar}}=0\, .\label{annex20}\fe}

When the symmetry orbit space $\ov\calXr$ is flat, this ansatz is evidently 
equivalent to a weak symmetry ansatz of the kind introduced by
Witten. The advantage of the prescription (\ref{annex20}) is that it is 
applicable even when $\ov\calXr$ is curved, and its disadvantage in that 
case is that it is not exactly applicable everywhere simultaneously,
 but only on the chosen cross section at  $t=0\, ,\ z=0$.
It can however be adopted as a very good approximation so long as
$\chir^{\!\Ar}$ is restricted to a range that is small compared with the
curvature scale of $\ov\calXr$.

The concrete implementation of such an approximation procedure is 
conveniently achievable, for the extended Witten models introduced 
above, by taking the local coordinates on the symmetry-orbit 
space $\ov\calXr$ to be specified by simply setting
{\be \chir^{\!_\Ar}=\frac{\Sigmar^\ard}{\Sigmar}\, ,\hskip 1 cm 
\Ar=\ard-1\, ,\hskip 1 cm \ard=2, ...,q+1\, .\label{annex21}\fe}
By substituting this in (\ref{annex6}), one obtains the $q$-spherical
metric (\ref{annex6}) in the explicit form given by
{\be {\rm d}\hat\Omegar^2=\delta_{_{\Ar\Br}}\, {\rm d}\chir^{\!_\Ar}\,
{\rm d}\chir^{\!_\Br} +\frac{(\delta_{_{\Ar\Br}}\, \chir^{\!_\Ar}\,
{\rm d}\chir^{\!_\Br})^2}{1-\chir^2}
\, .\label{annex22}\fe}
in which the deviation from flatness is attributable just to the last 
term, which will be negligible so long as the dimensionless quantity
{\be \chir^2=\delta_{_{\Ar\Br}}\chir^{\!_\Ar}\chir^{\!_\Br}
\label{annex23}\fe}
is very small compared with unity. 

In the minimally extended case, for which $q=2$, this metric will be 
that of an ordinary 2-sphere, with  coordinates expressible as
{\be  \chir^{\!_1}=  {\rm sin}\,\hat\thetar\, {\rm sin}\,\hat\varphir
\, , \hskip 1 cm\chir^{\!_2}= {\rm cos}\,\hat\thetar
\, .\label{annex24}\fe}
in terms of  spherical coordinates  of the usual kind, for which 
{\be {\rm d}\hat\Omegar^2=
{\rm d}\hat\thetar^2+ {\rm sin}^2\hat\thetar\, {\rm d}\hat\varphir^2
\, .\label{annex25}\fe}

Adoption of the convention that $ \chir^{\!_\Ar}= 0$, 
on the chosen initial cross section, as specified by  $t=0\, ,\ z=0$, 
is equivalent to requiring there that 
{\be \Sigmar^{_1}=\Sigmar, \hskip 1 cm \Sigmar^\ard =0\ \ \forall\ \ 
\ard\neq 1\, ,\label{annex26}\fe}
which implies by  (\ref{annex17})
that the first derivatives of $\Sigmar^{_1}$ will vanish there,
 {\be \dot\Sigmar{^{_1}} =0\, , \hskip 1 cm\Sigmar{^{\prime_1}}=0\, ,\fe}
and by (\ref{annex18}) that its second derivatives
will be given there by
{\be \frac {\ddot\Sigmar{^{_1}}}{\Sigmar}=-\delta_{_{\Ar\Br}}
\dot\chir{^{_\Ar}}\dot\chir{^{_\Br}}\, ,\hskip 0.6 cm
\frac {\dot\Sigmar{^{\prime_1}}}{\Sigmar}=-\delta_{_{\Ar\Br}}
\dot\chir{^{_\Ar}}\chir{^{\prime_\Br}}\, ,\hskip 0.6 cm
\frac {\Sigmar{^{\prime\prime_1}}}{\Sigmar}=-\delta_{_{\Ar\Br}}
\chir{^{\prime_\Ar}}\chir{^{\prime_\Br}}\, .\label{annex27}
\fe}
Since the metric (\ref{annex22}) has the form required for 
 the local geodicity
ansatz to take  the form (\ref{annex20}), whereby
the second derivatives of $\chir^{_\Ar}$  all vanish where
$\chir^{_\Ar}=0$, it follows that the second derivatives of
the other components of $\Sigmar^\ard$ will also vanish there,
{\be \ddot\Sigmar{^{\ard}}=\dot\Sigmar^\prime{^{\ard}}=
\Sigmar^{\prime\prime\,\ard}=0 \ \ \forall\ \ 
\ard\neq 1\, .\label{annex28}\fe}
We thereby obtain 
{\be \ddot\Sigmar{^{_1}}-\Sigmar{^{\prime\prime\,_1}}=\hat\wrd\,\Sigmar
\, ,\hskip 1 cm
 \ddot\Sigmar{^\ard}-\Sigmar{^{\prime\prime\,\ard}}=0 \ \ \forall\ \ 
\ard\neq 1\, ,\label{annex29}\fe}
where $\hat\wrd$ is defined according to (\ref{2_6}) as the
the trace of the target-space tensor 
$$\hat\wrd{^{_{\Ar\Br}}}=\chir{^{\prime_\Ar}}\chir{^{\prime_\Br}}-
\dot\chir{^{_\Ar}}\dot\chir{^{_\Br}}\, ,$$
so that, where $\chir^{_\Ar}=0$, it will be given, independently of 
$\varrhob$, by 
$$ \hat\wrd=\delta_{_{\Ar\Br}}(\chir{^{\prime_\Ar}}\chir{^{\prime_\Br}}
-\dot\chir{^{_\Ar}}\dot\chir{^{_\Br}})=
\frac{1}{\Sigmar^2}\delta_{\ard\brd}(\Sigmar{^{\prime\ard}}
\Sigmar{^{\prime\brd}}-\dot\Sigmar{^{\ard}}\dot\Sigmar{^{\brd}})\, .$$

Under these circumstances, the dynamical equations of the 
carrier subsystem (\ref{annex16}) will be satisfied automatically 
for $\ard\neq 1$, so this subsystem reduces to a just a single 
non trivial equation, which will take the form
{\be \frac{1}{\varrhob}\frac{{\rm d}}{{\rm d}\varrhob}\left(
\varrhob \frac{{\rm d}\Sigmar}{{\rm d}\varrhob}\right)
=\left(\hat\wrd +2\,\frac{\partial\hat\calVr}
{\partial(\Sigmar^2)}\right) \Sigmar\, .\label{annex30}\fe}
This  can be seen to be independent of the internal dimension $q$, 
and thus exactly the same as that of its analogue for the original 
unextended Witten model. It is evident that 
the carrier contribution to the kinetic part (\ref{annex_4a})
of the Lagrangian density (\ref{annex_4a})
will also reduce to the same form as for the original Witten model, 
namely
{\be \Lr_{\bf car}=-\frac{_1}{^2}\, \hat\wrd\,\Sigmar^2
\, .\label{annex31a}\fe}
 
Within the framework of the local geodicity ansatz, the preceeding 
work establishes that in terms of the constant parameter $\hat\wrd$ 
and the radially dependent field variable $\Sigmar^2$ the carrier
contribution in the extended model is indistinguishable
from the carrier contribution in the original model.
It follows that, in terms of the scalar trace $\hat\wrd$, the extended 
model will share, with the original Witten model, an equation of state 
of exactly the same simply harmonious form for the ensuing string model 
Lagrangian,
{\be \ov \Lr=\int\!2\pi\varrhob\, \Lr\,{\rm d}
\varrhob \, ,\label{annex32}\fe}
as obtained \cite{CP95,HC08} by integration over the cross section 
using field values satisfying the dynamical equations consisting of
 (\ref{annex30}) for the carrier amplitude in 
conjunction with the Higgs subsystem consisting of (\ref{annex11a})
and (\ref{annex11a}).


It is to be remarked that instead of interpreting the  independent
 variable $\hat\wrd$ in the simply harmonic equation of state 
as the trace  of the target space tensor $\hat\wrd{^{_{\Ar\Br}}}$,
it can equivalently be characterised as  the trace
{\be \ov\wgn=\ov\wgn_i^{\,\ i} \, ,\label{annex42a}\fe}
of the worldsheet tensor defined by (\ref{2_15}), which
in the local neighbourhood where $\ov\Aru_i{^{_\Ar}}$ vanishes
will take the simple form
$ \ov \wgn_{ij}=\hat\grd_{_{\Ar\Br}}\chir{^{\!_\Ar}_{\, ,\, i}}
\chir{^{\!_\Br}_{\, ,\, j}}$. As well as having the same trace,
{\be \ov\wgn=\hat\wrd\, ,\label{annex43}\fe} 
 the target-space tensor $\hat\wrd{^{_{\Ar\Br}}}$ and
base-space tensor $\ov\wgn_{ij}$ share their quadratic invariant
{\be \ov\wgn_i^{\,\ j}\ov\wgn_j^{\,\ i}=\hat\wrd{_{\!_\Ar}^{\ _\Br}}\,
\hat\wrd{_{\!_\Br}^{\ _\Ar}}\, ,\label{annex44}\fe}
which can thereby be seen to be positive definite, due to
the Euclidean signature of $\hat\grd_{_{\Ar\Br}}$
(whereas the signature of the 2-dimensional worldsheet metric 
$\ov\ggn_{ij}$ is Lorentzian).

\section{Conclusions}
\label{Annex3}

The upshot of the forgoing work is that the extension of Witten's
field model will have vortices macroscopically describable by a 
conducting string model of simply harmonic type, in the sense of 
being governed by a Lagrangian depending just on the trace 
$\ov \wgn$ of the base space tensor $\ov \wgn_{ij}$ defined by 
(\ref{2_15}) as the pullback of the prescribed target space metric 
$\hat\grd_{_{\Ar\Br}}$. It is instructive to present the properties 
of such models using a preferred worldsheet reference frame of 
orthonormal type, meaning
{\be \ov\ggn_{_{00}}=-1\, ,\hskip 1 cm \ov\ggn_{_{11}}=1\, ,
\hskip 1 cm \ov\ggn_{_{01}}=0\, ,\label{annex49}\fe}
that is chosen so as to diagiagonalise the pullback
$\ov \wgn_{ij}$, of which the non-vanishing components will be the
 eigenvalues that are given in terms of the  invariants 
$\ov\wgn$ and $\ov\wgn{_i^{\, \ j}}\ov\wgn{_j^{\, \ i}}$
by the expressions
{\be \ov\wgn_{_{00}}=\sqrt{\frac{\ov\wgn{^{\,2}}}{4}-
{\rm det}\{\ov\wgn\}\,}+  \frac{\ov\wgn}{2} \, ,\hskip 1 cm
\ov\wgn_{_{11}}=\sqrt{\frac{\ov\wgn{^{\,2}}}{4}
-{\rm det}\{\ov\wgn\}\,}- \frac{\ov\wgn}{2}\, ,\label{annex45}\fe}
in which the relevant  invariant determinant, namely that of 
$\ov\wgn_i^{\, \ j}$,  is defined as
{\be {\rm det}\{\ov\wgn\}=\ov\wgn{_{_0}^{\,\ _0}}
\ov\wgn{_{_1}^{\,\ _1}}-\ov\wgn{_{_0}^{\,\ _1}}
\ov\wgn{_{_1}^{\,\ _0}} =\frac{_1}{^2}(\ov\wgn{^{\,2}}-
\ov\wgn{_i^{\, \ j}}\ov\wgn{_j^{\, \ i})}\, .\label{annex46}\fe}
It is to be noticed that positive definite character of the target 
metric $\hat\grd_{_{\Ar\Br}}$ entails the non-negativity of the 
eigenvalues,
{\be \ov\wgn_{_{00}}\geq 0\, ,\hskip 1 cm \ov\wgn_{_{11}}\geq 0
\, \label{annex47},\fe}
and hence the non- positivity of the determinant: 
{\be\hat\wrd{_{\!_\Ar}^{\ _\Br}}\,\hat\wrd{_{\!_\Br}^{\ _\Ar}}
-\hat\wrd{^{\,2}}=-2\, {\rm det}\{\ov\wgn\}\geq 0
\, .\label{annex48}\fe}
It is also to be observed that the determinant will necessarily 
vanish -- meaning that the equality $\hat\wrd{_{\!_\Ar}^{\ _\Br}}
\,\hat\wrd{_{\!_\Br}^{\ _\Ar}}=\hat\wrd^{\, 2}$ 
will automatically be satisfied so that either $\ov\wgn_{_{00}}$
or $\ov\wgn_{_{11}}$ will be zero
-- in the special case \cite{LMMP09} for which the currents 
happen to be all aligned with the {\it  same} one-parameter 
subgroup generator in $\ov\calXr$, which is the only kind of
configuration for which a stationary circular vorton type
equilibrium state will be possible. 

In terms of the eigenvalues $\ov\wgn_{_{00}}$
and $ \ov\wgn_{_{11}}$, the corresponding 
non-vanishing components of the surface stress energy tensor 
(\ref{1_7}) will be given by the formulae
{\be \ov\Tr_{\!_{00}}=\kappar\,\ov\wgn_{_{00}}-\ov\Lr\, ,\hskip 1 cm 
\ov\Tr_{\,_{11}}=\kappar\,\ov\wgn_{_{11}}+\ov\Lr\, \label{annex50},\fe}
in which $\ov\Lr$ and $\kappar$ depend only on the trace,
$\ov\wgn=\ov\wgn_{_{11}}-\ov\wgn_{_{00}}$.

The original Witten field model gave rise to an ensuing conducting 
string model that was of ordinarily elastic type in the sense of 
having a target space dimension $q$ that was the same as the space 
dimension $p$ (not the spacetime dimension $p+1$) of the supporting 
worldsheet, which in the string case is simply $p=1$. It was 
found\cite{CP95} that a good approximation for the equation of state 
of the  ensuing string model could be provided in terms of a 
microscopic length scale $\delth_\star$ and a couple of mass scales 
$\mm$ and  $\mm_\star$ -- depending \cite{HC08} on the particular 
functional form of the interaction potential $\hat\calVr$ -- by the 
formula
{\be \ov\Lr=-\mm^2-\frac{_1}{^2}\mm_\star^{\,2}\, {\rm ln}\,
\{1+\delth_\star^{\,2}\,\ov\wgn\}\ \, ,\label{annex51}\fe}
which gives
{\be \kappar=\frac{\mm_\star^{\,2}\delth_\star^{\,2}}
{1+\delth_\star^{\,2}\,\ov\wgn}\, ,\hskip 1 cm \frac{1}{\kappar}
\frac{{\rm d}\kappar}{{\rm d}\ov\wgn}=\frac{-\delth_\star^{\,2}}
{1+\delth_\star^{\,2}\,\ov\wgn}\, .\label{annex52}\fe}
The work in the preceeding sections implies that the minimally 
extended Witten model with the same interaction potential
 $\hat\calVr$ will give rise to an ensuing conducting string 
model that will be governed by the {\it same} equation of state, 
even though it will not be of the ordinary elastic type with $q=p$, 
but of the hyperelastic type \cite{C07}
with $q=p+1$, that is to say with a target space having a dimension
that is equal to the space-time dimension of the supporting 
wordldsheet, namely $p+1=2$ in the string case under consideration. 
It can be seen from the characteristic equation (\ref{2_14}) that, 
according to (\ref{annex52}), the speed $\crd_{_{\rm L}}$ of longitudinal 
sound type perturbations relative to the preferred reference system 
 will be given in this hyperelastic case by
{\be \crd_{_{\rm L}}^{\,2}=\frac{1-\delth_\star^{\,2}(\ov\wgn_{_{00}}+
\ov\wgn_{_{11}})}{1+\delth_\star^{\,2}(\ov\wgn_{_{00}}+
\ov\wgn_{_{11}})}\, ,\label{annex53}\fe}
 where $\ov\wgn_{_{00}}$ and $\ov\wgn_{_{11}}$ are the -- necessarily
non-negative -- eigenvalues given by (\ref{annex45}), of which the 
sum  in (\ref{annex53}) will be expressible as $\ov\wgn_{_{00}}+
\ov\wgn_{_{11}}=\sqrt{ 2\,\hat\wrd{_{\!_\Ar}^{\ _\Br}}\,\hat
\wrd{_{\!_\Br}^{\ _\Ar}}-\hat\wrd{^{\,2}}\,}$. 

It is to be remarked that the dually related \cite{C89a,C95} 
``electric'' and ``magnetic'' varieties of the ordinarily 
elastic case obtained from the original unextended Witten model can 
be considered as degenerate limits of this hyperelastic case: 
the ``electric'' variety is characterised by 
$\ov \wgn=-\ov\wgn_{_{00}}$ with $\wgn_{_{11}}=0$, and the   
``magnetic'' variety is characterised by 
$\ov\wgn=\ov\wgn_{_{11}}$ with $\ov\wgn_{_{00}}=0$.

As well as the longitudinal modes characterised by (\ref{annex52}),
there will of course be extrinsic ``wiggle'' perturbation
of the world sheet, with propagation speed $\crd_{_{\rm E}}$,
which according to (\ref{1_6}) will be 
given  in terms of the energy density ${\calUr}=\Tr_{\!_{00}}$
and the string tension ${\calTr}=-\Tr_{\!_{11}}$
 by the universally \cite{C95,C01} valid formula
{\be \crd_{_{\rm E}}^{\,2}=\frac{\calTr}{\calUr}\, .
\label{annex54}\fe}
The novelty in the  minimally extended Witten model,
as contrasted with the original Witten model,
is that the ensuing string model has a third kind of perturbation
mode. As well as the longitudinaly polarised sound type modes 
governed by (\ref{annex53}) and the extrinsic ``wiggle'' modes 
governed by (\ref{annex54}), there will also be intrinsic
 ``shake'' modes characterised (with respect to the curved target 
space $\ov\calXr$) by the transverse polarization 
condition (\ref{2_11}) with propagation speed
$\crd_{_{\rm T}}$ given according to (\ref{2_12}) by
{\be \crd_{_{\rm T}}^{\,2}=1\, ,\label{annex55}\fe}
which simply means that (like the  perturbation modes
of the underlying field model with the flat target space
$\calXr$) these shear type modes will travel at the speed
of light.

Having completed the presentation of this simply harmonious 
hyperelastic string model, we still need to consider the extent 
to which its use is justifiable as a macroscopic string 
description of  vortex  defects in the minimal 
extension of the Witten field model. The foregoing derivation 
relied on a local geodicity ansatz whose applicability depended 
on a local flatness approximation whose range of validity is 
limited by the requirement that the dimensionless magnitude
$\chir^2$ defined by (\ref{annex23}) should remain small.

Subject to the conditions of (\ref{annex20}), the local geodicity
 ansatz means that for small but nonzero values of $t$ and $z$ 
the value of  $ \chir^{\!_\Ar}$ will be given approximately by 
$ \chir^{\!_\Ar}=\dot\chir{^{\!_\Ar}} t +\chir^{\!\prime_\Ar} z$, 
or more concisely, in terms of the worldsheet 
coordinates $\sigme^{_0}=t$ and $\sigme^{_1}= z$, by
$  \chir^{\!_\Ar}=\chir{^{\!_\Ar}_{\, ,\, i}} \,\sigme^{i}$. 
It follows that the required magnitude  $\chir^2$ will be given 
in this approximation by the formula
$  \chir^2=\ov\wgn_{ij}\,\sigme^{i}\,\sigme^{j}$.
Thus more particularly, with respect to the preferred longitudinal 
coordinate system characterised by (\ref{annex45}), 
it will  be given by
{\be \chir^2 =\ov\wgn_{_{00}}\,t^2+\ov\wgn_{_{11}}\, z^2
\, .\label{annex57}\fe}
 For the  approximate flatness approximation to be 
utilisable, the necessary condition,
{\be \chir^2 \ \ll \ 1\, ,\label{annex58}\fe}
must be satisfied in a range of $t$ and $z$ that is large
compared with the thickness $\deltag$ of the vortex deffect,
which entails the requirements $\ov\wgn_{_{00}}\,\deltag^2\ll 1$
and $\ov\wgn_{_{11}}\,\deltag^2\ll 1$. 

This leads us to the final conclusion, which is that the condition 
for negligibility of the curvature term in (\ref{annex22}) -- and 
thus for the validity of the simply harmonious conducting string 
model -- will be expressible simply as an upper bound on the 
quadratic field gradient invariant 
$\hat\wrd{_{\!_\Ar}^{\ _\Br}}\,\hat\wrd{_{\!_\Br}^{\ _\Ar}}$
(and hence by (\ref{annex48}) also on $\hat\wrd{^{\, 2}}$)
 in the form
{\be \hat\wrd{_{\!_\Ar}^{\ _\Br}}\,\hat\wrd{_{\!_\Br}^{\ _\Ar}}
\,\deltag^4 \ \ll\ 1\, .\label{annex59}\fe}
So long as the string thickness $\deltag$ is small enough, this 
condition will hold automatically by (\ref{annex48}) as a 
consequence of the requirement  for stability against longitudinal 
perturbations, namely the condition $\crd_{_{\rm L}}^{\,2}>0$, which 
according to (\ref{annex53}) will take the form
 {\be (2\,\hat\wrd{_{\!_\Ar}^{\ _\Br}}\,\hat\wrd{_{\!_\Br}^{\ _\Ar}}
- \hat\wrd{^{\,2}})\,\delth_\star^{\,4} <1\, .\label{annex60}\fe}
Within this allowed range there will clearly be no danger of
violation of (\ref{annex59}) provided that (as is usually supposed 
in the cosmological context envisaged by Witten) the length  scales 
associated with the carrier field are relatively large, and the 
mass scales correspondingly small, compared with those associated 
with the Higgs field:
{\be \delth_\star^{\,2}\gg \deltag^2\, , \hskip 1 cm \mm_\star^{\,2}
\ll \mm^2\, .\label{annex61}\fe}
There is never any possibility of instability of the shake modes 
characterised by (\ref{annex55}), and the usual extra requirement
  $\crd_{_{\rm E}}^{\,2}>0$,  for stability against extrinsic wiggle 
perturbations -- which is equivalent to the condition that  the 
tension $\calTr$ given by (\ref{annex50}) should be positive 
-- makes little difference in the circumstances of (\ref{annex61}), 
as it can be seen that it will hold automatically if (\ref{annex60}) 
is replaced by the only marginally stronger condition 
$ (2\hat\wrd{_{\!_\Ar}^{\ _\Br}}\,\hat\wrd{_{\!_\Br}^{\ _\Ar}}
- \hat\wrd{^{\,2}})\,\delth_\star^{\,4} <1- {\rm exp}\{-2\mm^2/
\mm_\star^{\,2}\}$.

\vfill\eject

\bigskip
{\bf Acknowledgements}
\medskip

 The author is grateful to Marc Lilley, Jerome Martin, and Patrick
 Peter for stimulating conversations.

\end{document}